# A New Market-Based Framework for Increasing Responsive Loads In Distribution Systems

Mohammad Panahazari[1], Minoo Mohebbifar[2], Vahid Nazari Farsani[1], Mahmoud-Reza Haghifam[1]
[1] Department of Electrical and Computer Engineering, Tarbiat Modares, University, Tehran, Iran
[2] Department of Electrical and Computer Engineering, Tabriz, University, Tabriz, Iran
*Emails: {m.panahazari, nazarivahid, haghifam}@modares.ac.ir, mohebbifar.minoo@tabriz.ac.ir*

*Abstract*—— Regarding the pervasive application of information and telecommunication technologies in the power distribution industry, responsive loads (RLs) have been widely employed in the operation of distribution and transmission systems. The utilization of these loads in the competitive environment of the power market has led to a decrease in costs and an increase in the flexibility of the distribution system and, consequently, the power system. This paper presents a framework for the competitive presence of RLs in local markets. The technical cooperation method of the Distribution System Operator (DSO) and Transmission System Operator (TSO), the persuasion mechanism of DSOs, and financial signals for getting and increasing the participation of consumers are represented based on local markets and market clearing mechanisms.

## NOMENCLATURE

A. Indices and Abbreviations

| | |
|---|---|
| $i$ | Index of system buses |
| $s$ | Index of DSOs |
| $r$ | Index of RLs |
| $G$ | Abbreviation of Generation buses |

B. Sets

| | |
|---|---|
| $N$ | Set of system buses |
| $T$ | Set of sub transmission substations |
| $L$ | Set of system branches |
| $R$ | Set of RLs |

C. Constants

| | |
|---|---|
| $C_{op}^{TSO}$ | TSO operation cost |
| $C_{RL}$ | Cost of using RLs |
| $C_{con}^{TSO}$ | Cost of using conventional methods |
| $C_i^G$ | Regulation bid for network generators at bus i |
| $C_{others}$ | Cost of other options in ancillary service market |
| $C_r^{RL}$ | Regulation bid for RL |
| $C_s^{DSO}$ | Regulation bid for DSOs |
| $B_l$ | Susceptance for branch L |
| $\Delta P_i$ | Unbalance on scheduled injected power in bus i |
| $P_l^{F.max}$ | Maximum power flow for branch l |
| $P^{G-}$ | Minimum regulation for generator |
| $P^{G+}$ | Maximum regulation for generator |
| $P_r^{RL-}$ | Minimum regulation for RL |
| $P_r^{RL+}$ | Maximum regulation for RL |
| $P_i^{DA}$ | Scheduled injected power at bus i |

D. Variables

| | |
|---|---|
| $P_i^{TD}$ | Injected real power from substation in bus i |
| $P_l^F$ | Power flow for branch l |
| $P_i^G$ | Regulated power for Generator in bus i |
| $P_s^{DSO}$ | Regulated power for DSOs |
| $\delta_i$ | Voltage angle at bus i |
| $P_r^{RL}$ | Regulated power for responsive load at bus i |

## I. INTRODUCTION

With the advent of new technologies, advanced communication, and telecommunication devices, the possibility of using RLs has been raised. RLs are dispatchable, flexible, available, and observable loads with fast responsiveness [1]. As a matter of fact, a vast majority of consumers could be categorized as RLs if the requirements of the demand side and network are fulfilled. Most loads fall into this category, including electric vehicles (EVs), smart homes equipped with smart metering devices and controllable appliances, industrial and commercial loads, and energy storage systems (ESS).

RLs are able to participate in various operation and management programs, including congestion management [2]. In addition, it is possible for RLs to participate in the day-ahead electricity market in order to prevent the occurrence of price spikes [3]. Besides, RLs can be beneficial to facilitate the conventional procedure of unit commitment problems [4]. Other benefits of using these loads in the network are an increase in energy efficiency and a decrease in environmental pollution [5]. Moreover, distribution systems can also benefit from RLs [6]; DSOs can use RLs for controlling voltage, decreasing loss, improving reliability, and managing congestion in the network [7-9]. Table I. shows the application of RLs from DSO and TSO points of view.

Since there are a high number of RLs in the network, their direct participation in the day-ahead market and ancillary service market is impractical. Therefore, the participation of aggregators in the market is crucial for reducing the number of market players and decreasing uncertainties by balancing and smoothing the participation rate of RLs. Independent entities, DSOs, and retailers can play aggregators' role in the market. Some effort has been made by researchers to investigate TSO-DSO coordination with a focus on power system flexibility. Authors in [10] evaluate different mechanisms for the flexibility market. A method for flexibility range estimation at TSO-DSO boundary nodes presented in [11] and [12] proposes a decentralized restoration method with coordination on the transmission and distribution system. Furthermore, uncertainties such as renewable energy volatile generation, wholesale prices, and the behavior of EV owners can significantly affect the available flexibility resources to TSOs. However, the great performance of deep learning-based methods in power system problems, including motor control and fault detection [13] [14], has motivated some authors to Manage Distributed Flexibility uncertainties using Deep Learning [15].

Deployment of RLs requires telecommunication infrastructures and encouraging financial mechanisms. One of the essential technologies for integrating RLs is the Advanced Metering Infrastructure (AMI) system[16], which can be costly to deploy. Consequently, DSOs often expand this system only when they can realize a return on investment. It is expected that this benefit along with the contribution of RLs in distribution and transmission network management increase the number of RLs in the network. In other words, the financial competition between DSOs leads to their investment in the expansion of AMI and ICT technologies which are essential for RLs' growth. Section II of this paper proposes a new framework to create an environment in which there is financial competition between DSOs and also suggests a new method of interaction between DSOs and TSO in order to increase RLs and subsequently make the network more flexible. In section III the proposed framework is used for modeling a real-time balancing market. Later in section IV, the simulation of the model is presented and the paper arrives at conclusions in section V.

## II. Proposed Framework

The growth in the number and capacity of RLs in the network, along with the development of their preparation infrastructure in the management and control of the network, leads to an increase in the network's flexibility. This section introduces a framework to boost flexibility by the creation of competition between aggregators. In this paper, DSOs are considered RL aggregators motivated to invest in the development of the required ICT basis of RLs by not only financial incentives but also technical benefits that guarantee an improvement in network controllability and manageability. Consequently, this investment results in an increase in RLs' number and their application in the network.

TABLE I. BENEFITS OF RLS

| DSO | TSO |
|---|---|
| Congestion Management | Congestion Management |
| Improving System Reliability | Loss Reduction |
| Voltage Regulation | Improving System Resiliency |
| Loss Reduction | Improving System Reliability |
| Increasing Renewable Resources Hosting Capacity | Reducing Operational Costs |
| Increasing System Resiliency | Increasing System marginal Stability |
| Outage Management | Frequency Control |
| | Increasing System Security |
| | Decreasing Power Shortage |

There is a TSO→DSO→Load relation in every power system that shows a control and power flow direction. When RLs are not present, the TSO sends controlling and dispatching commands to DSOs, who perform various activities such as load shifting between feeders, adjusting the network structure, switching capacitors, and even load shedding during emergencies. In order to enhance the flexibility of the distribution network, a competitive mechanism needs to be devised, enabling DSOs to derive benefits from supporting TSO. So, there would be an increase in the probability of DSOs investing in RLs and telecommunication infrastructures. Therefore, the quantity of RLs and their contribution is raised, and the result is a reduction in operation costs and an increment in flexibility rate and the satisfaction level of customers. Fig. 1. shows the conceptual framework consisting of information and control directions.

As discussed in section I, TSO can use RLs for a wide range of goals. In each of them, TSO sends its request to the central market in which DSOs participate. Upon receiving the request, DSOs transfer it to their local market and execute their programs, which vary according to different requests and DSOs or aggregators. In a distribution system, a DSO can either work with predetermined fixed contracts with a central controlling system to execute requests or run a local market and ask the consumers to send their bids. After checking them, DSOs send aggregated bids to the central market. After receiving bids from DSOs, TSO chooses the best suggestions regarding operation cost and technical constraints and announces the winners. There are three information flows based on the communication-telecommunication system. The first flow is sending contribution requests from TSO to the central market and DSOs to consumers for a specific program. The second flow of information is receiving and gathering consumers' bids using aggregators, and the last one is a controlling flow containing the best bids that TSO sends to DSOs to be executed.

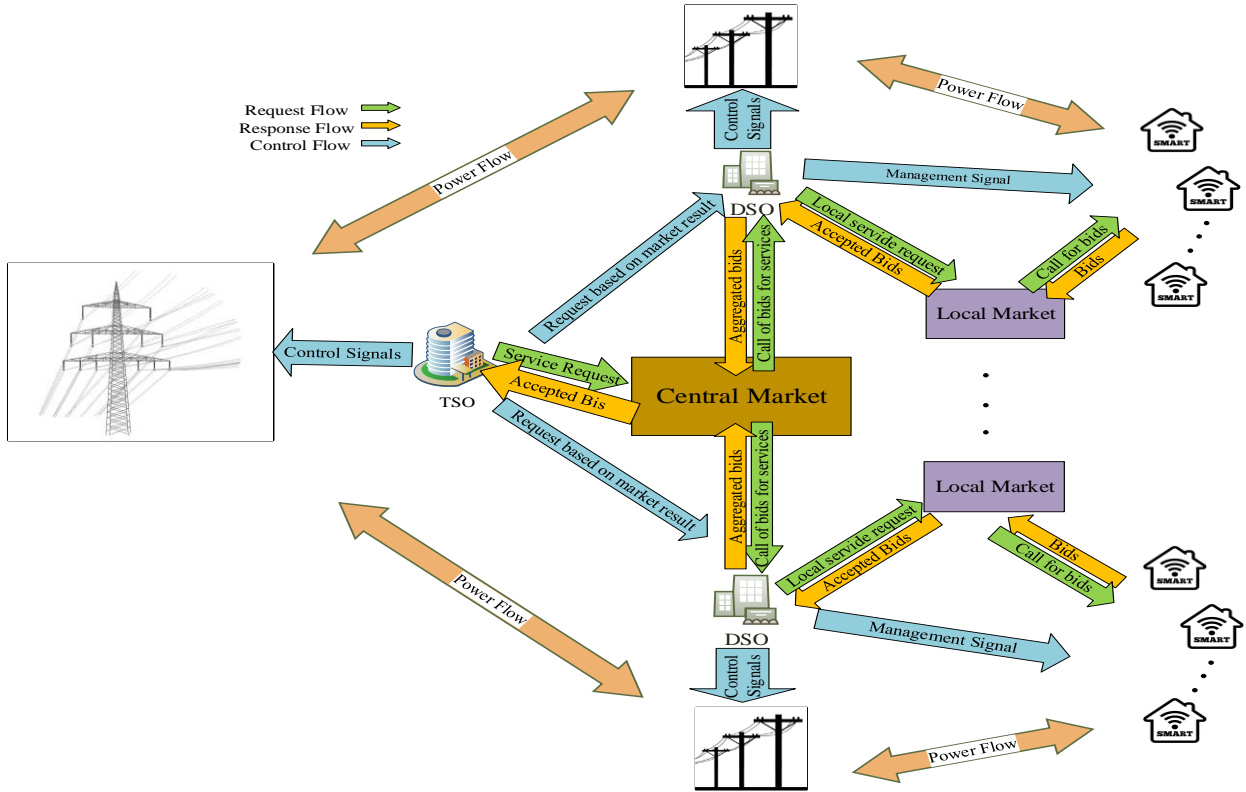

Figure 1. Proposed framework for incresing RLs

The method TSO chooses to utilize the framework depends on the infrastructure and its intelligence. In general, a real-time telecommunication platform is needed to send requests to the market, run central and local markets, and aggregate bids in a short time. Currently, SCADA systems have provided the connection between TSO and DSOs in power systems, and the only requirement of the proposed framework is providing an ICT platform in the distribution system intending to run local markets. Therefore, AMIs play the most crucial role in communicating with consumers and DSOs, which is essential for congestion management, loss reduction, and load shedding in the network. However, executing other programs like frequency control and ancillary services with the help of RLs needs a more advanced infrastructure, like advanced control systems that can be implemented in a central or distributed manner. All in all, using new technologies provides a situation to have a flexible distribution system with RLs. Smart homes equipped with Home Energy Management System (HEMS) can help establish the proposed framework [17] introduced in this paper.

It is necessary to express the importance of the proposed framework. Considering the progression in communication technologies and the tendency to build a competitive environment to reduce costs, moving towards the proposed framework is inevitable, and the future power system will work based on this framework.

### III. FORMULATION OF THE PROPOSED FRAMEWORK

TSO has many tools to manage the network. Fig. 2. illustrates some of these tools. As a tool, TSO can use RLs for load management in all network buses. There are several distribution systems and DSOs in every network bus volunteer to participate in TSO's requested programs. One of the buses in this condition is shown in Fig. 3. TSOs choose the suggestions with the lowest costs because, from their point of view, the difference between DSOs lies solely in the costs of their suggestions. The operation cost of TSO in a management operation like balancing the network is

$$C_{opr}^{TSO} = \min\{C_{RL} + C_{Con}\} \quad (1)$$

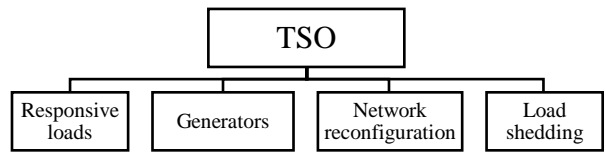

Figure 2. Some TSO's management tools

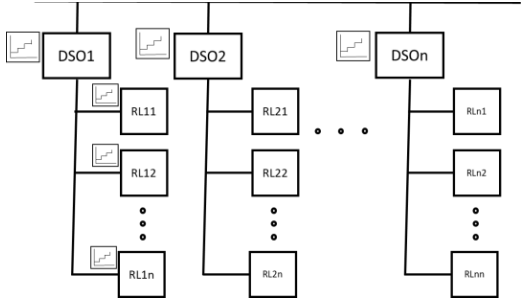

Figure 3. One of network buses

According to Eq. 1. TSO uses RLs if they have a lower cost than the other management options, considering the technical constraints of the network. As shown in Fig. 3. after the announcement of TSO's need in the central market, RLs send their bids to DSOs in every local market. Then, DSOs choose and aggregate the bids of their subareas with their specific methods. Naturally, DSOs should take into account the amount of their benefits from aggregating RLs to help the execution of TSO's programs. In fact, the benefits can be divided into two groups: technical and financial. The former is about improving distribution system operational conditions like loss reduction, voltage level improvement, congestion management, etc. The latter and the more motivating benefit is the financial one. The more RLs DSOs gather, the more benefit they will gain.

An ancillary service market is considered to check the functionality of the proposed network. Assume that the day-ahead power market has been run, the results have been presented, and the operator has scheduled generation based on the forecasts of renewable generations. However, the real generation is different from the predictions. In this situation, TSO decides to run a real-time market to reschedule the generation and wants to use RLs to balance the network. In the proposed framework, TSO sends the request to the central market, and DSOs receive this request. After that, DSOs start running the local markets to gather bids from RLs.

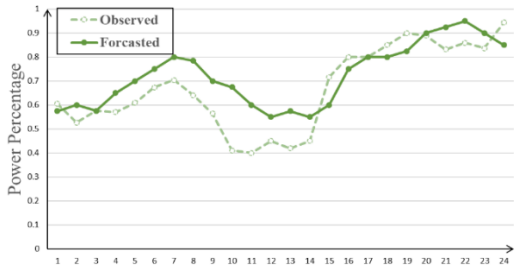

Figure 4. Forcasted and Observed Power of wind power plant

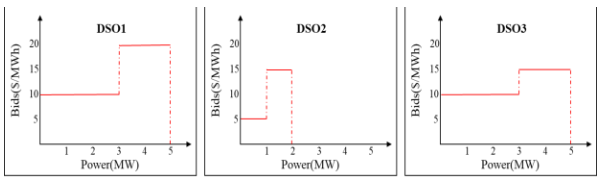

Figure 5. DSOs' aggregated bids in central market

Based on aforementioned discussions, cost function of TSO in the real time market on the proposed framework is

$$\min\left\{\sum_{i\in G_0} C_i^G P_i^G + \sum_{s=1}^{S} C_s^{DSO} P_s^{DSO} + C_{others}\right\}. \quad (2)$$

Eq. 2 depicts balancing cost of the network using different tools such as using the present controllable generators which are located in transmission system, RLs of distribution system or any other way. Transmission constraints are [18]:

**TABLE II.** PARTICIPANTS BIDS IN REAL TIME BALAMCING MARKET

| ID | Name | Subsystem | Offer (MW) | Price($/MWh) |
|---|---|---|---|---|
| 1 | G1 | T | -5 | 25 |
| 2 | D11 | D1 | 2 | 20 |
| 3 | D12 | D1 | 2 | 10 |
| 4 | D13 | D1 | 1 | 10 |
| 5 | D21 | D2 | 1 | 15 |
| 6 | D22 | D2 | 1 | 5 |
| 7 | D31 | D3 | 2 | 10 |
| 8 | D32 | D3 | 2 | 15 |
| 9 | D33 | D3 | 1 | 10 |

**TABLE III.** RLS LOCATION IN DISTRIBUTION NETWORK

| DSO | loads location number |
|---|---|
| 1 | 634, 645, 675 |
| 2 | 634, 680, 611 |
| 3 | 632, 671, 611 |

$$P_i = \sum_{j\in N_0}(B_{ij}(\delta_i - \delta_j)) \quad \forall i \in N_0 \quad (3)$$

$$P_i^{DA} + P_i^G + \Delta P_i = P_i \quad \forall i \in N_0 - N_0^{TD} \quad (4)$$

$$P_i^{DA} + P_i^G + \Delta P_i = P_i + P_i^{TD} \quad \forall i \in N_0^{TD} \quad (5)$$

$$P_i^{DA} + P_i^G + \Delta P_i = P_i + P_i^{TD} \quad \forall i \in N_0^{TD} \quad (6)$$

$$P_i^{G-} \le P_i^G \le P_i^{G+} \quad \forall i \in G_0 \quad (7)$$

$$P_l^F = (B_{ij}(\delta_i - \delta_j)) \quad \forall l = (i,j) \in L \quad (8)$$

$$-P_l^{F,\max} \le P_l^F \le P_l^{F,\max} \quad \forall l \in L. \quad (9)$$

Interconnection point of transmission and distribution system constraints are [13]

$$P_t^F = P_i^{TD} = -P_j^{TD} \quad \forall t = (i,j) \in T \quad (10)$$

$$-P_t^{F,\max} \le P_t^G \le P_t^{F,\max} \quad \forall t \in T. \quad (11)$$

Cost function of DSO in local market follows Eq. 12.

$$\min \sum_{s\in S} C_r^{LR} P_r^{LR} + C_{loss} \quad (12)$$

The aim of a DSO is to choose the suggestions with lower costs in order to win the competition against the other DSOs in the central market. On the other hand, network losses incur costs for the DSO. Therefore, including these losses in the cost function of the DSO in the local market can help to make a more accurate decision. All RLs have maximum and minimum redispatchable power constraints.

$$P_r^{RL-} \leq P_r^{RL} \leq P_r^{RL+} \qquad \forall i \in R \qquad (13)$$

## IV. SIMULATION

In this section, a real-time balancing market is simulated for the 5-bus IEEE test system [19]. It is assumed that a wind farm is located at bus 2. The corresponding predicted power and the real observed power are presented in Fig. 4. Three DSOs are considered in this system in a configuration that DSO1 and DSO2 are located at bus 4, and DSO3 is at bus 3. There is also a controllable generator in bus 1. In addition, the suggestions of RLs and controllable generators are shown in Table II. For simplicity, we use the IEEE 13-node test system to model distribution networks [20] and locate the RLs on the nodes listed in Table III. We also use the simplified loss calculation model from [21]. A DSO aims to increase the contribution of RLs to decrease the cost of its network loss.

Moreover, DSO is awarded in proportion to the amount of power it can provide for regulation in the central market. Hence, the strategy of DSO is presenting suggestions with lower prices considering its network loss cost (Eq. 12). Having this strategy, DSO presents a stepped bid to participate in the central market (Eq. 2), which is the aggregation of RLs' bids located in its subarea. It is clear that the local market would be run under a condition of mismatch between observed and forecasted power. So, according to Fig. 4 DSOs do not participate in the central market at hours 3 and 17. Suggestions of DSOs to the central market are shown in Fig. 5. The results of running the central balancing market by TSO are presented in Fig. 6. Because the only suggestion of generation reduction is from the power plant, in the whole time when generation reduction is needed the power plant is the winner. Additionally, since the suggested price of the DSO2 is low, when load reduction is needed, DSO2 is usually selected by the central market for regulation. Obviously, it earns more profit than the other DSOs. Fig. 7. Shows the regulated power of RLs in all local markets.

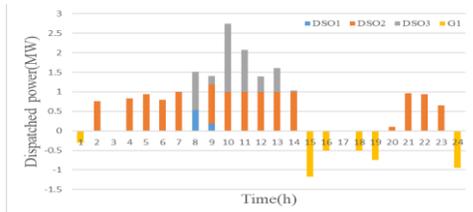

Figure 6. Active power dispatch of central balancing market participants

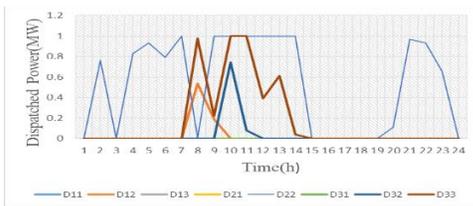

Figure 7. Dedicated power for RLs in local markets

## V. CONCLUSION

In this paper, a market-based framework is proposed to create a competitive environment for DSOs with the goal of increasing RLs in distribution systems and it is shown that a rise in the number of RLs increases the flexibility of the network. In addition, the way of using local markets in order to gather suggestions of RLs is introduced and different strategies of DSOs in local markets are discussed. Furthermore, a real time balancing market in the proposed framework is formulated and a sample network is also used to check the functionality of the framework. Consequently, it is presented that using the proposed framework can be effective in increasing the contribution of RLs and the motivation of DSOs. Also, DSOs can improve technical conditions of their systems in the case of using the proposed framework.